\documentclass[12pt]{iopart}
\usepackage{iopams}
\usepackage{graphicx}
\usepackage{lineno}%

\usepackage[margin=0pt]{caption}

\begin{document}
\title[Enhanced beam transport via space charge mitigation...]{Enhanced beam transport via space charge mitigation in a multistage accelerator for fusion plasma diagnostics}

\author{M. Nishiura$^{1, 2}$, K. Nakamura$^{2}$, K. Ueda$^{1}$, A. Shimizu$^{1}$, H. Takubo$^{1}$, M. Kanda$^{1}$, T. Ido$^{3}$, M. Okamura$^{4}$}

\address{$^{1}$National Institute for Fusion Science, 322-6 Oroshi-Cho, Toki 509-5292, Japan\\
$^{2}$Graduate School of Frontier Sciences, The University of Tokyo, Kashiwa, Chiba 277-8561, Japan\\
$^{3}$Department of Advanced Energy Engineering, Kyushu University, Kasuga, Fukuoka 816-8580, Japan\\
$^{4}$Collider-Accelerator Department, Brookhaven National Laboratory, Upton, New York 11973, USA
}
\ead{nishiura@nifs.ac.jp}
\vspace{10pt}
\begin{indented}
\item[]July 2025
\end{indented}

\begin{abstract}
Efficient transport of high-current negative ion beams is critical for accurate plasma potential diagnostics using heavy-ion beam probe (HIBP) systems in magnetically confined fusion plasmas. However, strong space-charge effects often degrade transport efficiency, particularly for heavy ions such as Au$^-$. In this study, we demonstrate a substantial improvement in beam transport by introducing an electrostatic lens effect through optimized voltage allocation in a multistage acceleration system. Numerical simulations using IGUN, supported by experiments with the LHD-HIBP system, show that this approach effectively suppresses space-charge-induced beam divergence and loss. Without requiring mechanical modifications to the beamline, the optimized configuration enables a 2–3 fold increase in Au$^-$ beam current injected into the tandem accelerator. Consequently, plasma potential measurements were extended to higher-density plasmas, reaching line-averaged electron densities up to $1.75 \times 10^{19}$~m$^{-3}$ with improved signal-to-noise ratio. This technique offers a compact, practical, and highly effective solution for transporting high-current heavy-ion beams under space-charge-dominated conditions. Beyond its impact on plasma diagnostics, the method is broadly applicable to a wide range of accelerator systems, including those used in scientific and industrial applications where high-intensity beam transport is required.
\end{abstract}

%
\submitto{\NF}
%
\maketitle
%
%

\section{Introduction}

Advancing fusion power generation requires the development of both plasma physics understanding and diagnostic instrumentation. Among various diagnostics, the heavy ion beam probe (HIBP) has been developed to measure plasma electric potential and its fluctuations, which are closely related to confinement performance through the radial electric field. In the Large Helical Device (LHD), a strong magnetic field of approximately 3 T necessitates the use of MeV-class heavy ion beams to access the plasma core.

The LHD-HIBP system employs a tandem accelerator capable of applying a nominal terminal voltage of 3 MV to produce the Au$^+$ probe beams with energies up to 6 MeV. The primary beams are injected into the plasma, where collisions with electrons and ions ionize them to the secondary beams of Au$^{2+}$. The resulting secondary beam passes through the plasma. An energy analyzer allows the evaluation of the plasma potential from the energy difference from the initial beam~\cite{Ido_2010, shimizu_2010}.

Zonal flows have been observed in CHS~\cite{Fujisawa_2004}, and potential fluctuations associated with GAMs and Alfvén eigenmodes have been reported in the T-10 and TJ-II tokamaks~\cite{Melnikov_2016}. The HIBP system in the LHD has been successfully used for measurements of radial electric fields during impurity transport~\cite{Ido_2010}, energetic particle-driven geodesic acoustic modes (EGAM)~\cite{Ido_2016}, and particle transport studies~\cite{Nishiura_2024}. In LHD deuterium experiments~\cite{Takeiri_2017}, the threshold power for the formation of an electron internal transport barrier is lower in deuterium plasmas than in hydrogen plasmas, accompanied by a clear potential transition observed with HIBP~\cite{Kobayashi_2022}.

To enhance the performance of the LHD-HIBP system, a plasma sputter-type negative ion source was developed to improve the injection beam current and has been successfully utilized for many years. Subsequently, to validate the ionization efficiency from Au$^-$ to Au$^+$ in a gas cell of the tandem accelerator, the relevant ionization cross sections were theoretically derived and evaluated both experimentally and theoretically. The measured conversion efficiencies were obtained with the theoretical predictions~\cite{Nishiura_2008, Nishiura_2008_NIFS_Rep}. It was also clarified that the dominant ionization processes differ depending on the beam energy: in the keV range, electron impact ionization prevails, while in the MeV range, ionization due to collisions with protons and highly charged ions becomes dominant. This energy dependence of ionization mechanisms is a distinguishing feature of the LHD-HIBP system, setting it apart from conventional HIBPs used in smaller devices.

In this study, we report the implementation of a new cesium sputter-type negative ion source that improves the signal-to-noise ratio (S/N) and enables the use of multiple sputtering targets. The ion source was first characterized on a test stand and then installed in the LHD-HIBP system. However, the beam current measured at the entrance of the tandem accelerator was significantly lower than that obtained on the test stand, indicating substantial beam losses along the low-energy beam transport line.

These results highlight that simply increasing the extracted negative ion beam current is insufficient to improve the injected or transmitted beam current in the tandem accelerator. Therefore, it is essential to identify the cause of these transport losses and to develop a solution to mitigate them.

To address this issue, we investigated the origin of the beam loss and developed a method to improve the transport efficiency on the injection side of the tandem accelerator. Specifically, an electrostatic lens effect was introduced by optimizing the electrode voltage allocation in the multistage accelerator tube. Both numerical simulations and beam transport experiments confirmed that this approach effectively mitigates beam divergence caused by space-charge effects, particularly for low-energy, high-current heavy ion beams. As a result, the negative ion beam current injected into the tandem accelerator was significantly increased without requiring any mechanical modifications to the beamline. Using the optimized beam transport configuration, plasma potential measurements in the LHD were successfully demonstrated.

\section{HIBP system for plasma diagnostics in the LHD}
\subsection{Negative ion beam extraction on test stand}

The cesium sputter-type negative ion source (High Voltage Engineering, Model 860C Multi Sample Source) produces cesium ions by thermionic emission from a heated filament. These ions bombard a metal target, producing negative ions via sputtering. The source is equipped with a revolver-type holder capable of loading up to ten metal targets, each embedded with Au or Cu and used as a sputtering material. This ion source has previously demonstrated an Au$^-$ beam output of up to 290~$\mu$A at Brookhaven National Laboratory~\cite{Thieberger_1983}.

Prior to integration into the LHD-HIBP system, the beam extraction characteristics of the ion source were evaluated at a dedicated test stand. The distance from the extraction electrode of the ion source to the Faraday cup used for these measurements was 591~mm, matching the corresponding distance to FC0 in the LHD-HIBP system. Figure~\ref{fig:comparison_of_beam_currents} shows the dependence of the extracted negative ion beam current on the extraction voltage for the Model 860C source. The Einzel lens located at the exit of the ion source was tuned to maximize the beam current measured at the Faraday cup. While a direct comparison with the previously used plasma sputter-type negative ion source~\cite{Nishiura_2008} is not straightforward due to differences in beam energy and operational conditions, comparable performance was observed, with Au$^-$ and Cu$^-$ beam currents reaching approximately 65~$\mu$A and 135~$\mu$A, respectively. Following confirmation of stable operation and favorable beam extraction characteristics on the test stand, the negative ion source was installed on the tandem accelerator for implementation in the LHD-HIBP system.

\renewcommand{\figurename}{Figure}

\begin{figure}[ht]
\centering
\captionsetup{margin=0pt}
\includegraphics[width=1\linewidth]{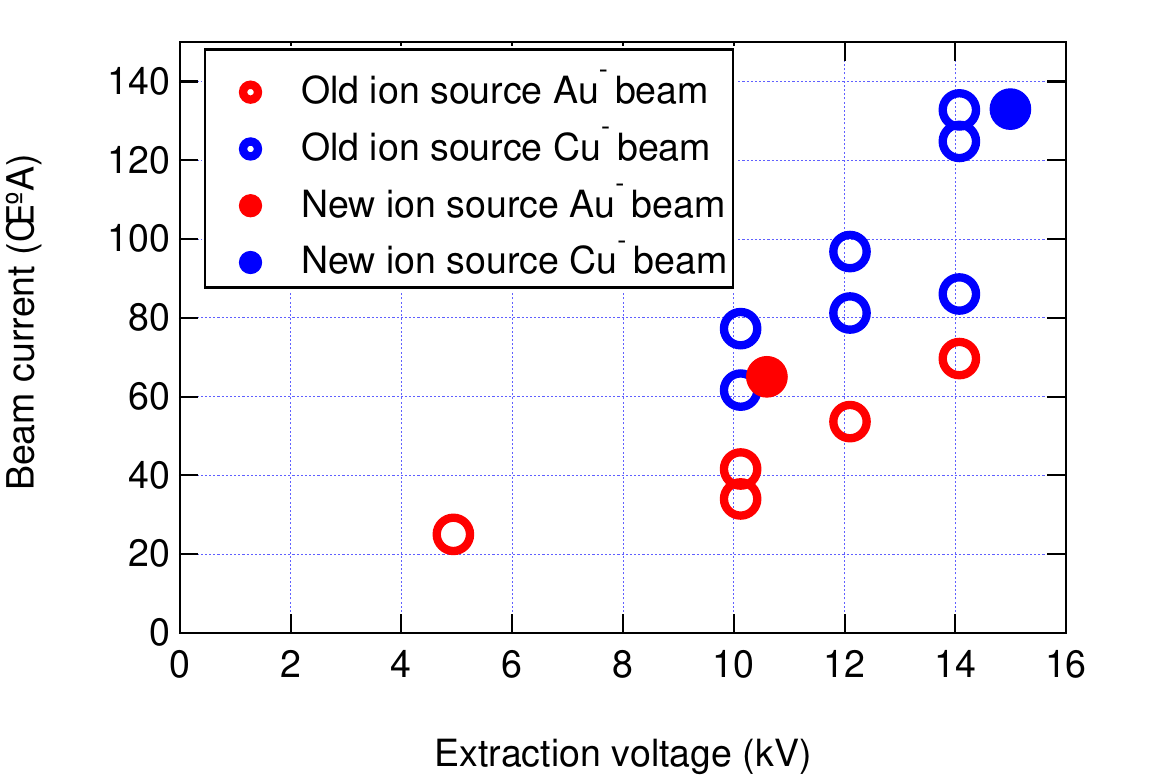}
\caption{Comparison of negative ion beam currents between the old plasma sputter type and the new cesium sputter type negative ion sources at the test stand. Open circles: old negative ion source \cite{Nishiura_2008}, closed circles: new negative ion source(this study).}
\label{fig:comparison_of_beam_currents}
\end{figure}

\subsection{Low energy beam transport of a tandem accelerator for LHD-HIBP system}
The Large Helical Device (LHD) is a large-scale superconducting toroidal fusion device equipped with two continuous helical coils with a poloidal pitch of $l=2/m=10$. It produces a toroidal magnetic field of $B_t = 2.75$~T and has major and minor plasma radii of $R = 3.9$~m and $a = 0.6$~m, respectively. The LHD is capable of steady-state operation via external superconducting coils and benefits from the inherent characteristics of the heliotron configuration, which eliminates the need for current drive and is intrinsically free from disruptions~\cite{Motojima:LHD:1999}.

Figure~\ref{fig:overview_of_HIBP}(a) shows an overview of the LHD, the injection-side beamline of the LHD-HIBP tandem accelerator, and the energy analyzer. The structure of the multistage accelerator tube is illustrated in Figure~\ref{fig:overview_of_HIBP}(b), while its cross-sectional schematic is presented in Figure~\ref{fig:overview_of_HIBP}(c). The multistage accelerator consists of four electrodes, with adjacent electrodes connected by 335~M$\Omega$ resistors. The voltages applied to electrodes 1, 2, 3, and 4 are denoted as V1, V2, V3, and VG, respectively. A total acceleration voltage of 44--48~kV is applied between V1 and VG, equally divided such that $\mathrm{V1 - V2 = V2 - V3 = V3 - VG}$. The energy of the negative ion beam extracted from the ion source is set to 16--20~keV, and the beam energy at the injection into the tandem accelerator is fixed at 64~keV.

Au$^-$ beams extracted at approximately 21~keV from the negative ion source are focused by an Einzel lens and subsequently accelerated to 64~keV using a multistage accelerator tube. The beam is then directed through a sector magnet, which removes impurity ions and allows the selective injection of Au$^-$ or Cu$^-$ beams into the tandem accelerator. A critical issue encountered was that the beam current measured at the entrance of the tandem accelerator did not reproduce the performance observed on the test stand. Specifically, although a beam current of 100~$\mu$A was obtained at FC0 (located at the exit of the ion source), the current measured at FC1 (positioned just upstream of the tandem accelerator) was only 20~$\mu$A, indicating significant beam losses along the transport path. The beamline between FC0 and FC1 includes both the multistage accelerator tube and the sector magnet. Attempts to mitigate the losses by adjusting the Einzel lens focus downstream of the ion source did not restore the beam current to the test stand level. These results suggest that the observed losses are more likely to be attributed to beam divergence caused by lens position, focal length, or space-charge effects, rather than to insufficient focusing or geometric misalignment.

\begin{figure*}[t]
\makebox[10cm][l]{(a)LHD and HIBP diagnostic system.}\\
\centering
\includegraphics[clip, width=5.0in]{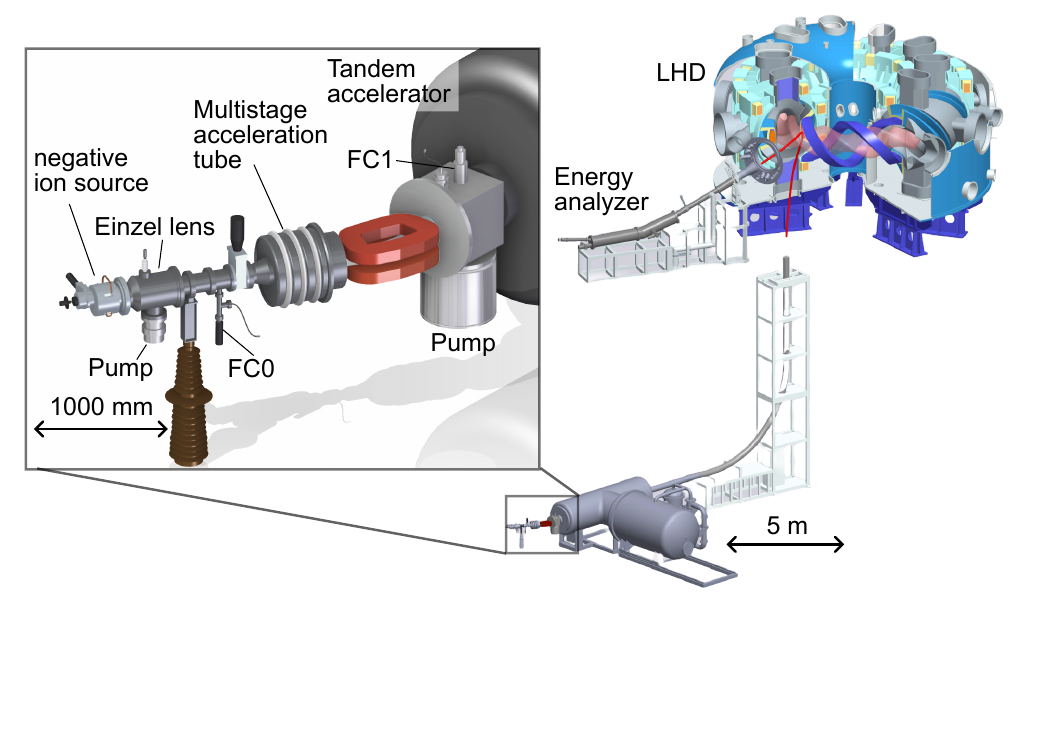}
    \vspace{-15pt}
    \centering
    \begin{minipage}{0.4\linewidth}
        \vspace{-25pt}
        \centering
        \includegraphics[width=\linewidth]{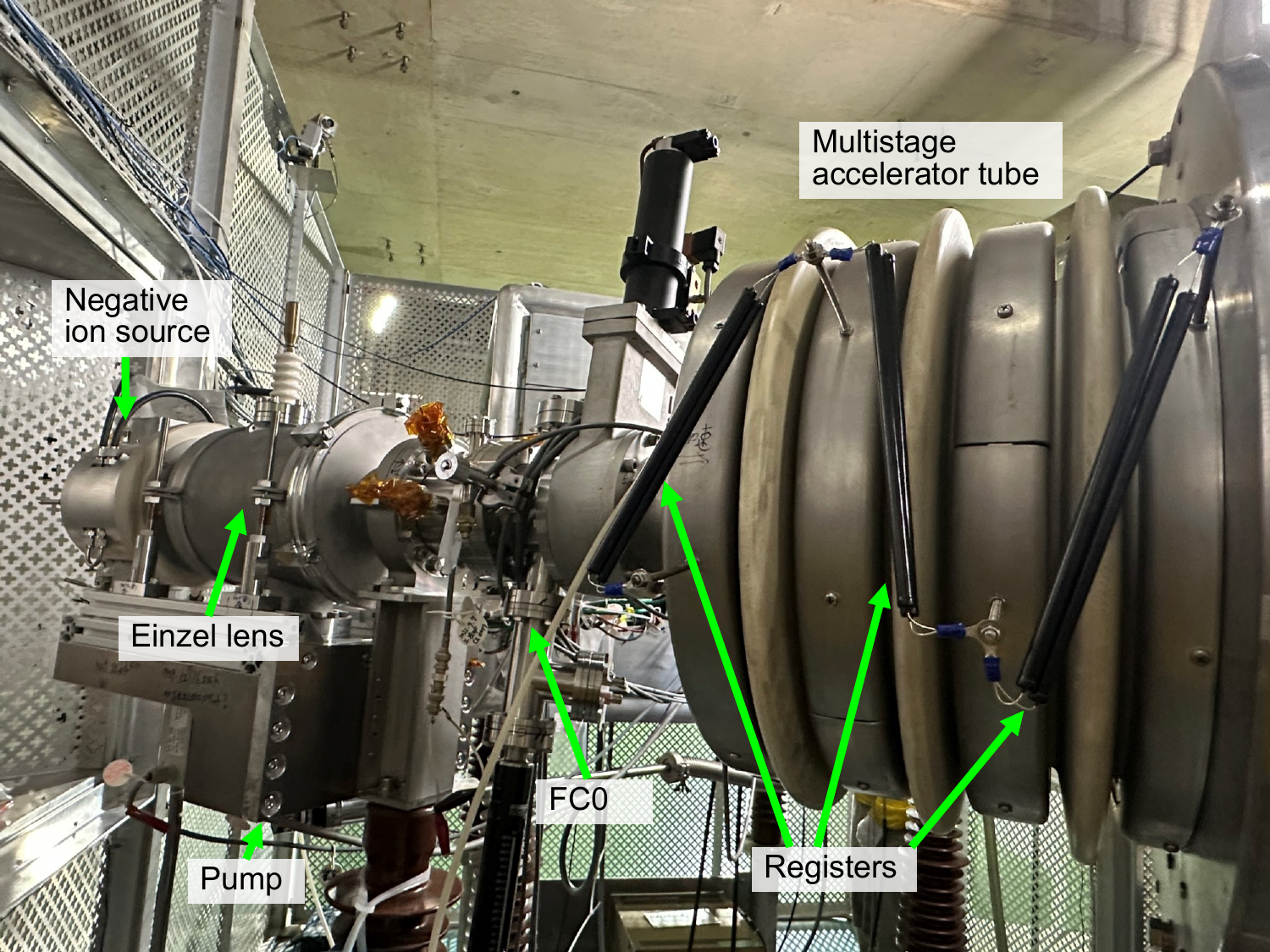}
        \textbf\textmd{{(b) Low energy beam transport from the negative ion\\ source to the multistage accelerator tube.}}
    \end{minipage}
    \hfill
    \begin{minipage}{0.5\linewidth}
        \vspace{-25pt}
        \centering
        \includegraphics[width=\linewidth]{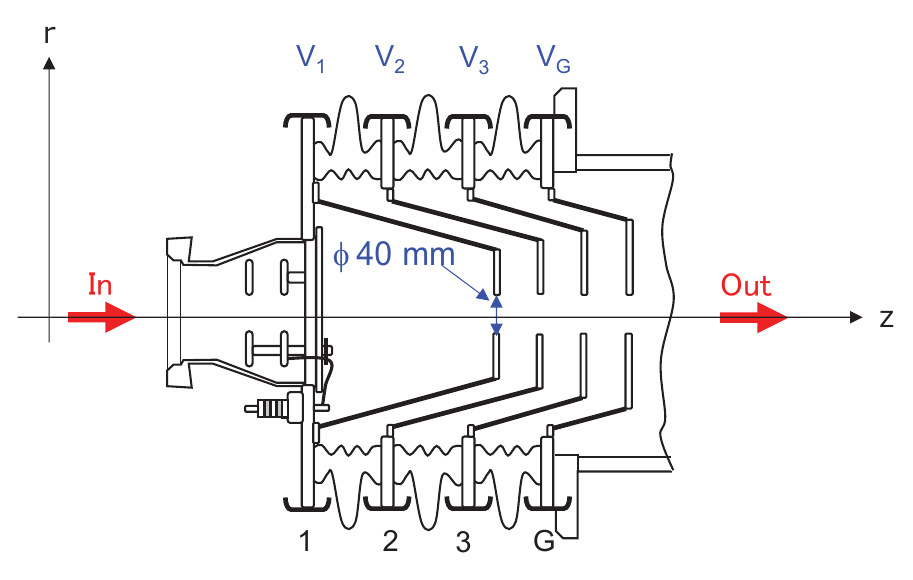}
        \textbf\textmd{{(c) Schematic view of the multistage accelerator tube.}}
    \end{minipage}
    \vspace{15pt}
\setcounter{figure}{1}

\caption{(a) Schematic of the LHD-HIBP system (negative ion source, tandem accelerator, beam transport line, and energy analyzer) and LHD. Enlarged view of the low-energy beamline upstream from the tandem accelerator [10]. FC0: $\phi$30 mm slit for beam profile measurement with a linear actuator, FC1: $\phi$30 mm Faraday cup.}
\label{fig:overview_of_HIBP}
\end{figure*}

Modifying the beamline components or the sector magnet poles in the LHD-HIBP system is technically challenging and impractical. To improve beam transmission efficiency, it was essential for the beam to traverse the narrowest section of the beam duct located within the sector magnet without significant loss. One potential solution involved introducing an electrostatic lens at the entrance of the sector magnet to reduce beam divergence. However, spatial constraints prevented the installation of such a component. As an alternative, we investigated the feasibility of incorporating an electrostatic lens effect into the existing multistage accelerator tube by optimizing the voltage allocation across its electrodes.

\begin{figure}
    \centering
    \includegraphics[width=0.8\linewidth]{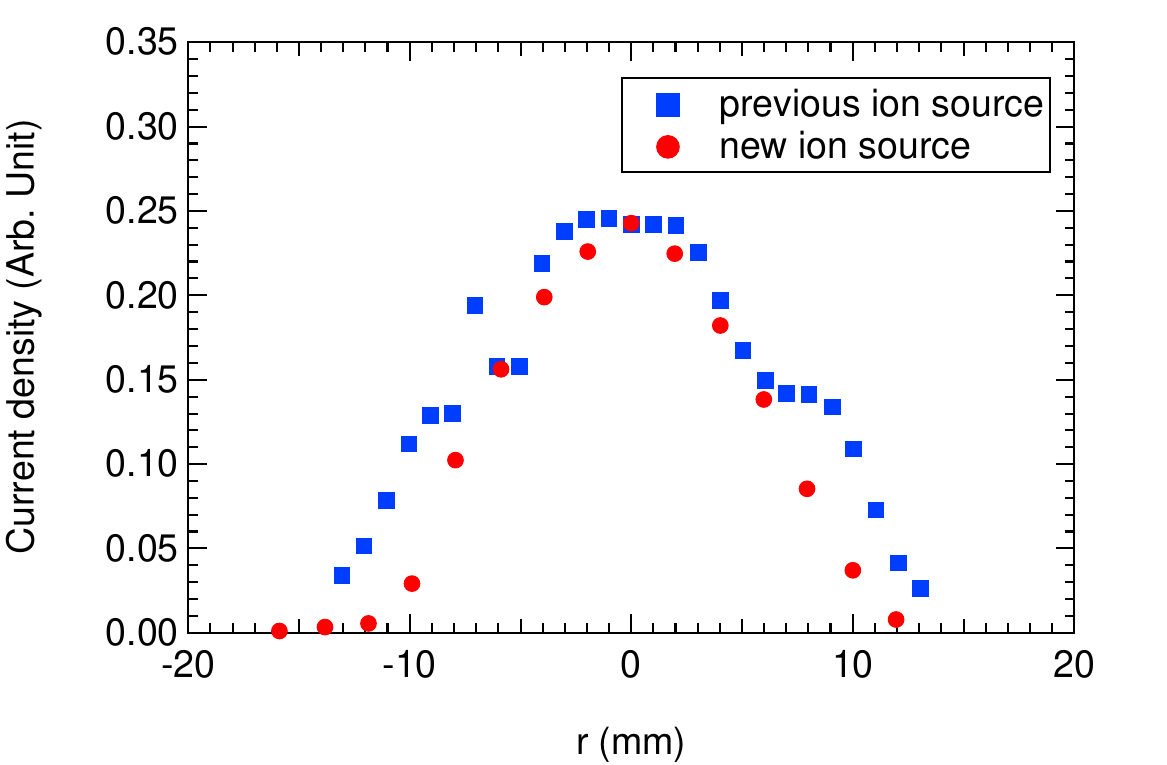}
    \caption{Comparison of Au$^-$ beam current profiles at FC0 between the previous and new negative ion sources. The previous negative ion source and new ion source had a beam energy of 13.9 keV and 21 keV, respectively.}
    \label{fig:beam_profiles}
\end{figure}

The voltages applied to each electrode of the multistage accelerator tube can be adjusted to both accelerate and focus the negative ion beam. To determine the optimal voltage configuration, the input beam current and divergence angle at the entrance of the accelerator tube are required for trajectory calculations. Beam extraction experiments were performed to characterize the beam profile and divergence at FC0. Figure~\ref{fig:beam_profiles} shows the Au$^-$ beam profiles measured using the Faraday cup FC0 (with an entrance slit diameter of $\phi = 30$~mm), located at $z = 530$~mm, where $z = 0$~mm corresponds to the position of the sputter target in the negative ion source. The beam current profile was obtained by vertically scanning the Faraday cup. The full width at half maximum (FWHM) of the Au$^-$ beam profile was found to be 24~mm, which is nearly identical to the beam diameter observed with the plasma sputter-type ion source used before~\cite{Nishiura_2008}.

\subsection{Simulation of low energy beam transport by IGUN simulation}
\begin{figure*}
  \centering
  \includegraphics[width=0.8\textwidth]{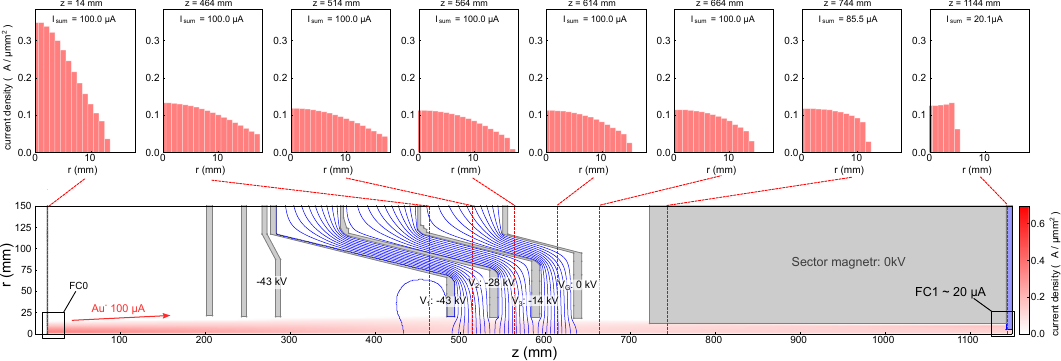}
  \caption{Model for the beam trajectory calculation and the result of the Au$^-$ beam trajectories. The voltages V1, V2 and V3 for a multistage accelerator tube, modeled from FC0 to FC1, z = 0 mm: FC0, z = 489 mm: V1 at electrode 1, z = 539 mm: V2 at electrode 2, z = 589 mm: V3 at electrode 3, z = 638 mm: VG to electrode 4 through the aperture of 40 mm in diameter. The electrode voltage is equally applied with $\mathrm{V1- V2 = V2- V3 = V3- VG}$. The equipotential lines are in blue. The current density of Au$^-$ beam is in red. The upper panel shows the Au$^-$ beam profile at each position.}
  \label{fig:model and ion trajectory}
\end{figure*}

\begin{figure*}
  \centering
  \includegraphics[width=1.0\textwidth]{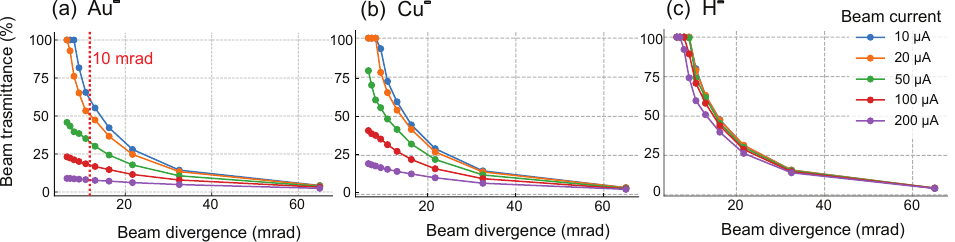}
  \caption{Dependence of negative ion beam transmittance calculated on initial divergence angle and beam current. (a) Au$^-$ beam, (b) Cu$^-$ beam, and (c) H$^-$ beam with an initial negative ion beam energy of 21 keV. The voltages for the multistage accelerator tube are applied equally to the electrodes.}
  \label{fig:dependence of negative ion beam transmittance}
\end{figure*}

The beam transport code IGUN~\cite{Patel:IGUN:2008} was employed to simulate the transport of ion beams from FC0 to FC1, located just upstream of the entrance to the tandem accelerator. Although IGUN was originally developed for positive ion beam simulations, it can be adapted for negative ion beams by reversing the signs of the applied voltages. For the present study, this adaptation enabled accurate trajectory calculations for low-energy, high-current negative ion beams under axisymmetric conditions. A cylindrical coordinate system was adopted, with the beam propagation direction defined along the $z$-axis. The Faraday cup FC0, positioned downstream of the ion source, was set as the origin ($z = 0$~mm), and FC1, located immediately before the tandem accelerator, was placed at $z = 1144$~mm. The multistage accelerator tube and the beam duct of the sector magnet were positioned along this axis between FC0 and FC1. The initial beam energy of the Au$^-$ ions was set to 21~keV. The multistage accelerator tube was operated with a total acceleration voltage of $-43$~kV, distributed equally across the electrodes such that $\mathrm{V1 - V2 = V2 - V3 = V3 - VG}$. The initial beam profile used in the simulation was based on the experimentally measured data shown in figure~\ref{fig:beam_profiles}.

To reproduce the experimentally observed beam currents of 100~$\mu$A at FC0 and 20~$\mu$A at FC1, the divergence angle of the Au$^-$ beam was numerically scanned. From this analysis, the initial divergence angle was determined to be approximately 10~mrad. These values were subsequently used as input parameters for IGUN simulations. The resulting beam trajectories for the Au$^-$ beam are shown in figure~\ref{fig:model and ion trajectory}. The simulation results revealed that significant beam losses occur within the electrodes of the multistage accelerator tube, the beam duct section of the sector magnet used for mass separation, and the aperture at FC1. To further clarify the origin of these losses, the dependence of beam transmission on both the initial divergence angle and the beam current was systematically investigated. The current was varied from 10 to 100~$\mu$A to examine the impact of space-charge effects. Figures~\ref{fig:dependence of negative ion beam transmittance}(a)--(c) show the transmission efficiency as a function of the initial divergence angle for Au$^-$, Cu$^-$, and H$^-$ beams, respectively. For all negative ion species, transmission efficiency decreases markedly with increasing divergence angle. In particular, heavier ion species exhibit a more pronounced drop in transmission at beam currents exceeding 20~$\mu$A, indicating that space-charge-induced beam divergence plays a dominant role in beam losses for high-mass ions. As shown in figure~\ref{fig:dependence of negative ion beam transmittance}(a), the transmission of a 100~$\mu$A Au$^-$ beam drops to approximately 20\% at a divergence angle of 10~mrad. This result suggests that, even with improved ion source performance and higher beam currents, it is intrinsically difficult to transmit more than 20~$\mu$A without addressing space-charge effects. Therefore, we explored whether beam losses could be mitigated by optimizing the electrode voltage distribution within the multistage accelerator tube.

\begin{figure}
    \centering
    \begin{minipage}{0.7\linewidth}
        \makebox[\linewidth][c]{(a) Au$^-$}
        \vspace{-10pt}
        \includegraphics[width=\linewidth]{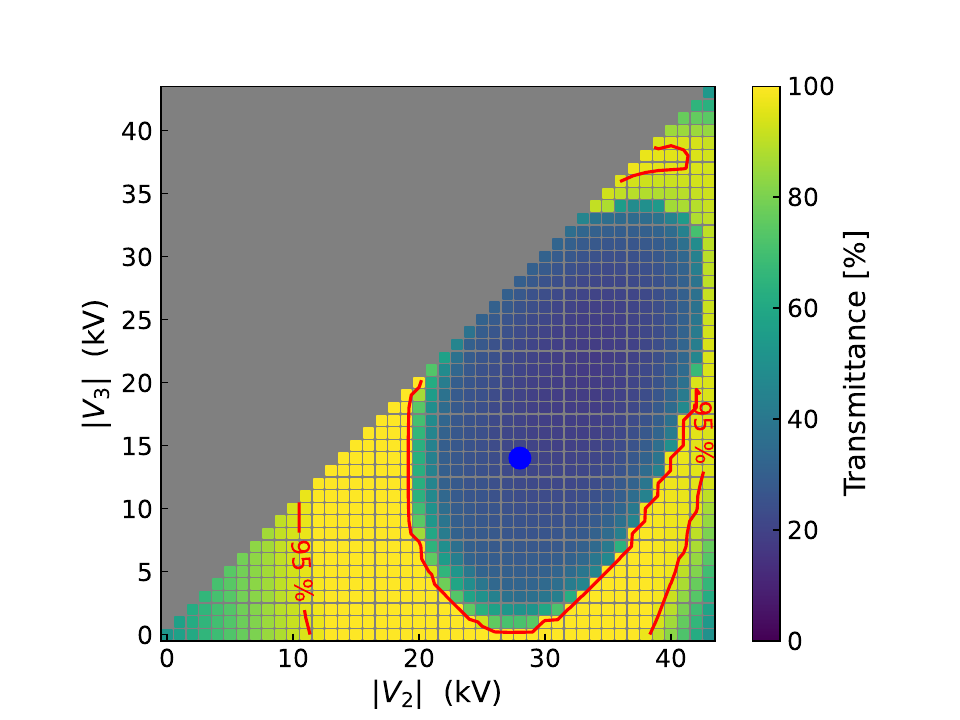}
        \label{fig:transmittance_Au}
        \vspace{15pt}
    \end{minipage}
    \hfill
    \begin{minipage}{0.7\linewidth}
        \makebox[\linewidth][c]{(b) Cu$^-$}
        \vspace{-10pt}
        \includegraphics[width=\linewidth]{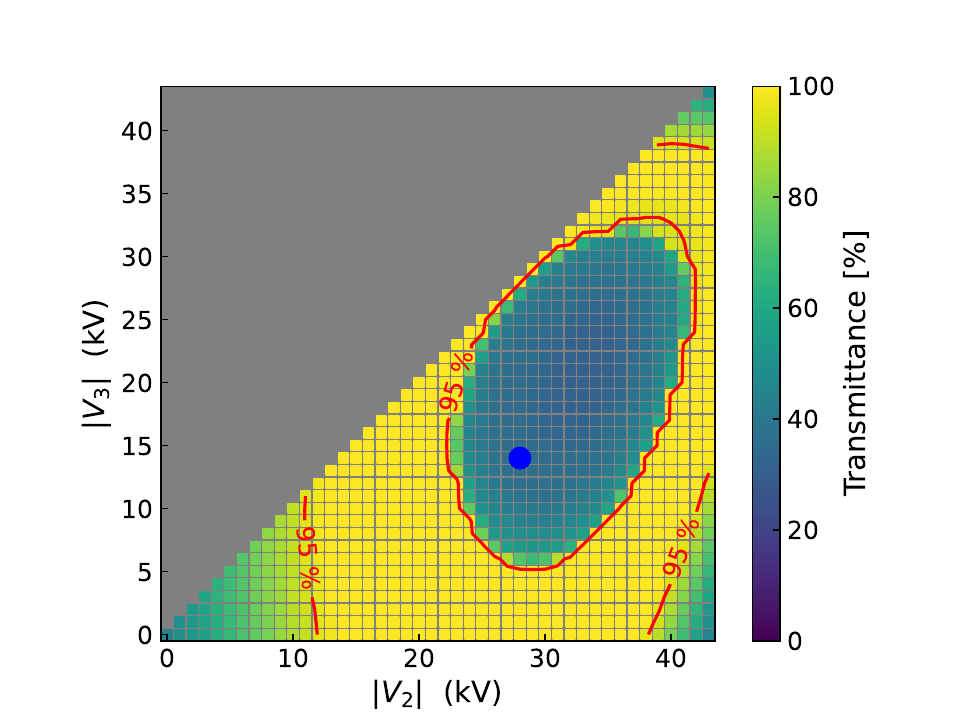}
        \label{fig:transmittance_Cu}
    \end{minipage}
    \caption{Maps of ion beam transmittance for electrode voltages V2 and V3 of multistage accelerator tube. Closed circles indicate the setting of the equal voltage between electrodes.}
    \label{fig:ion_beam_transmittance_maps}
\end{figure}

Figure~\ref{fig:ion_beam_transmittance_maps}(a) and (b) show transmittance maps for Au$^-$ and Cu$^-$ beams, respectively, under a beam current of 100~$\mu$A and an initial divergence angle of 10~mrad. The voltage of the first electrode was fixed at V1 = $-43$~kV, while the voltages V2 and V3 were systematically varied. The results indicate that the conventional equal-voltage configuration—where voltages are evenly divided by resistive dividers—corresponds to a region of low transmittance. In contrast, optimized voltage settings for V2 and V3 significantly enhance beam transmission, achieving efficiencies of 95\% or higher upon crossing a specific threshold. A comparison of the transmittance regions reveals that the Au$^-$ beam exhibits a narrower high-transmission window than the Cu$^-$ beam. This is attributed to the stronger space-charge effects associated with the higher mass of Au$^-$ at a given energy. These results demonstrate that space-charge-induced beam losses can be effectively mitigated by tuning the electrode voltage distribution in the multistage accelerator tube.

\subsection{Experimental optimization of beam transport in multistage accelerator tube}
\begin{figure}
    \centering
    \begin{minipage}{0.75\linewidth}
        \makebox[\linewidth][c]{(a) Au$^-$}
        \vspace{-10pt}
        \centering
        \includegraphics[width=\linewidth]{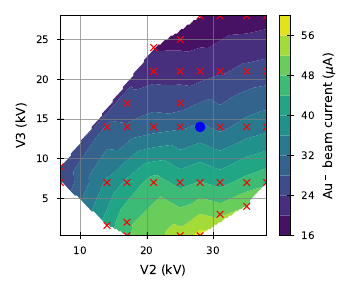}
        \label{fig:exp:Au}
    \end{minipage}
    \hfill
    \begin{minipage}{0.75\linewidth}
        \makebox[\linewidth][c]{(b) Cu$^-$}
        \vspace{-10pt}
        \centering
        \includegraphics[width=\linewidth]{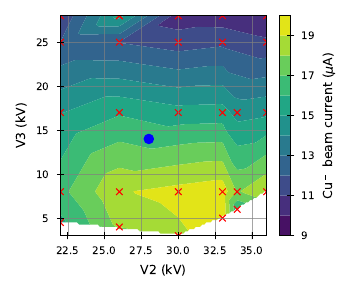}
        \label{fig:exp:Cu}
    \end{minipage}
    \caption{Maps of the negative ion beam current at FC1 on electrode voltages V2 and V3 of the multistage accelerator tube. (a)  Au$^-$ beam, and (b) Cu$^-$ beam. The cross symbols are measured values by FC1. The contours are interpolated by using measured beam currents. Closed circles indicate the setting of the equal voltage between electrodes. }
    \label{fig:ion_beam_transmittance_maps_experiment}
\end{figure}

Experiments were conducted to validate the improvement in negative ion beam transport efficiency through the electrostatic lens effect provided by the multistage accelerator tube, as demonstrated in the previous section. For this purpose, the resistors initially connected between the multistage accelerator electrodes were removed, and independent variable high-voltage power supplies were connected to the second and third electrodes (V2 and V3). The energy of the Au$^-$ beam injected into the multistage accelerator tube was set to 17~keV, and the total voltage difference between the first and fourth electrodes (V1 and VG) was fixed at 47~kV. The measured transmittance of the Au$^-$ and Cu$^-$ beams as a function of V2 and V3 is shown in figure~\ref{fig:ion_beam_transmittance_maps_experiment}. The operational parameters of the negative ion source were held constant during the experiments: the sputtering target power supply was set to $-6$~kV and 1.13~mA, and the ionizer power supply to 9.81~V and 16.92~A. The extraction voltage was $-15.7$~kV, the Au$^-$ beam energy was 21.7~keV, and the Einzel lens voltage downstream of the ion source was set to $-17.6$~kV.

For the Au$^-$ beam, the maximum transmitted beam current was observed at V2 = 28~kV and V3 = 0.3~kV. Under conventional equal-voltage settings, the beam current at FC1 was $-32.7~\mu$A, whereas it increased to $-57.5~\mu$A when the optimized voltages were applied—representing an improvement by a factor of approximately 1.76. Further increases in beam current may be possible if V3 is swept to more negative values. In the case of the Cu$^-$ beam, the highest beam current was obtained at V2 = 33~kV and V3 = 5~kV, where the beam current increased from $-15.4~\mu$A to $-19.8~\mu$A, corresponding to an enhancement factor of approximately 1.3. Notably, the optimal voltages for Cu$^-$ beams lie in the same general direction on the transmittance map as those for Au$^-$, i.e., increasing V2 and decreasing V3 relative to the conventional equal-voltage configuration.

\begin{figure}
    \centering
    \includegraphics[width=0.8\linewidth]{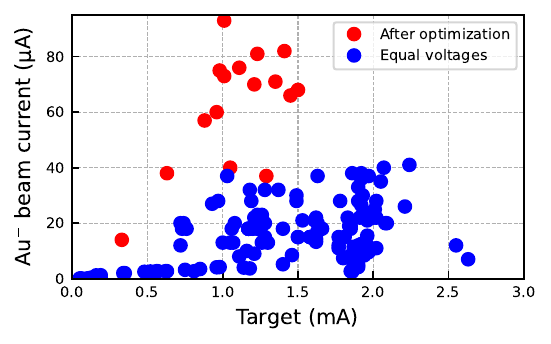}
    \caption{Characteristics of Au$^-$ beam current at FC1 versus target current of the negative ion source before and after the voltage optimization of the multistage accelerator tube. It can be seen that the negative ion beams are transported efficiently.}
    \label{fig:characteristics of Au at FC1}
\end{figure}

Figure~\ref{fig:characteristics of Au at FC1} presents a comparison of the Au$^-$ beam current measured at FC1 before and after the optimization of the multistage accelerator voltages during LHD plasma experiments. Since direct measurements at FC0 were not possible during these experiments, the Au target current in the negative ion source was used as a proxy for the negative ion beam current. Generally, for a fixed cesium supply condition, a higher target current corresponds to increased production of Au$^-$ ions, and thus a higher extracted beam current. This correlation was observed even prior to optimization. Following the voltage optimization, the beam current at FC1 increased by a factor of approximately two—from 38~$\mu$A to a maximum of 81~$\mu$A—as the target current at 1.0~mA. As a result of the optimization, the upper limit of the target current—corresponding to the output beam current of the negative ion source—was restricted to 1.5~mA. This limitation arose because the power supply capacity of the downstream beam transport system, extending from the tandem accelerator to the LHD injection port, was insufficient to accommodate the increased load associated with higher negative ion beam currents.

\begin{figure}
    \centering
    \includegraphics[width=0.8\linewidth]{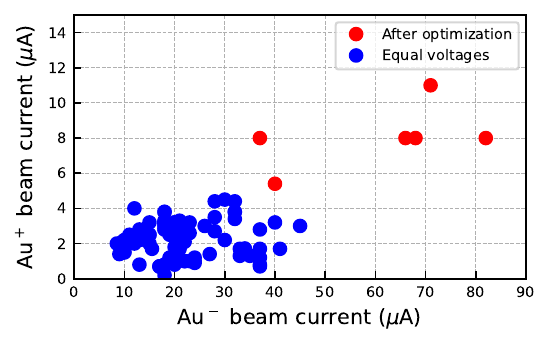}
    \caption{Output Au$^+$ beam current relative to the injection Au$^-$ beam current before and after passing through the tandem accelerator is compared before and after optimizing the voltages of the multistage accelerator tube.}
    \label{fig:Au beam current after tandem accerelator}
\end{figure}

Figure~\ref{fig:Au beam current after tandem accerelator} shows the relationship between the injected Au$^-$ beam current and the resulting Au$^+$ beam current after acceleration through the 3~MV tandem accelerator. Under the conventional equal-voltage configuration of the multistage accelerator tube, the injected Au$^-$ beam ranged from 10 to 40~$\mu$A, while the corresponding Au$^+$ beam current was limited to 3–4~$\mu$A. After optimizing the electrode voltages, the injected Au$^-$ beam increased to 38–81~$\mu$A, and the resulting Au$^+$ beam current improved significantly to 6–11~$\mu$A.

\begin{figure}
    \centering
    \begin{minipage}{0.85\linewidth}
        \makebox[\linewidth][c]{(a)}
        \vspace{-10pt}
        \centering
        \includegraphics[width=\linewidth]{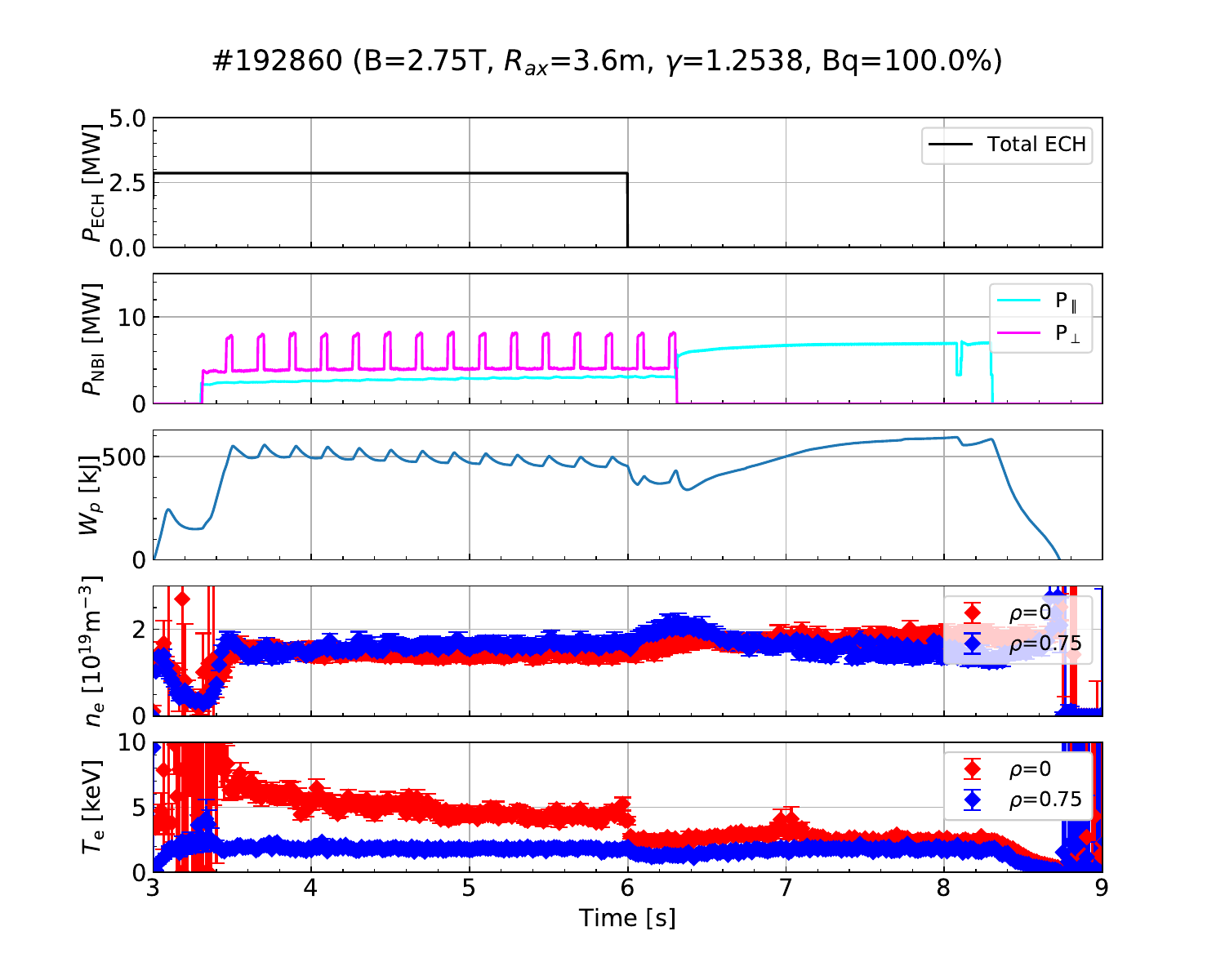}
        \label{fig:summary_LHD192860}
    \end{minipage}
    \hfill
    \begin{minipage}{0.8\linewidth}
        \makebox[\linewidth][c]{(b)}
        \vspace{-10pt}
        \centering
        \includegraphics[width=\linewidth]{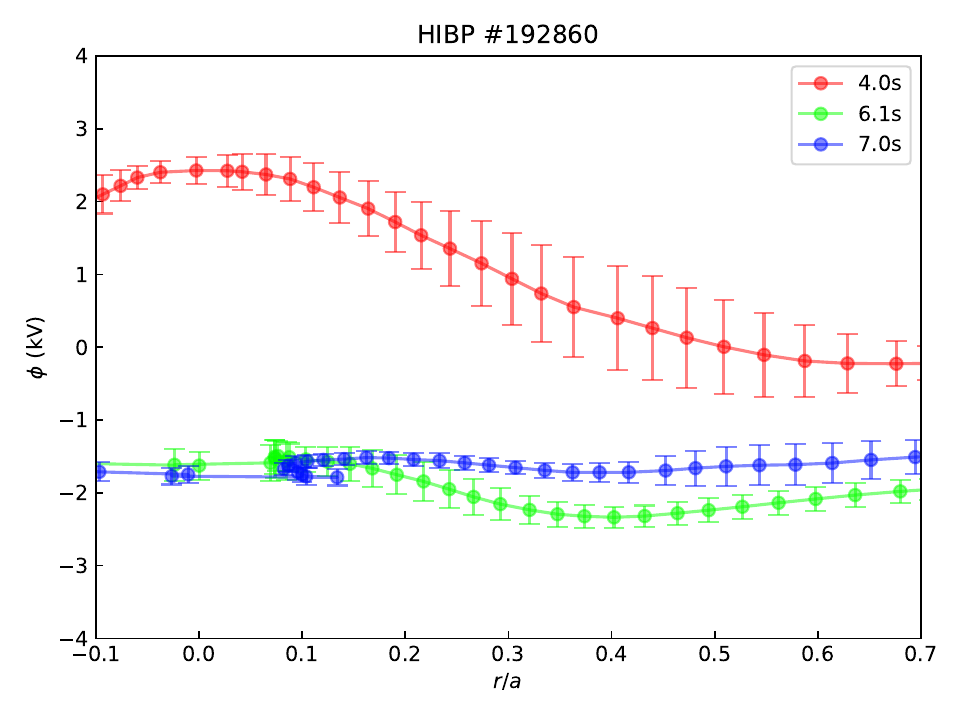}
        \label{fig:potential_profiles}
    \end{minipage}
    \caption{Discharge waveforms and plasma potential profile measurement in LHD. (a) LHD shot\#192860 for heating powers of ECH and parallel NBIs(beam energies are 137-177 keV) and perpendicular NBIs(beam energy is 40 keV), diamagnetic energy, and electron densities and temperatures at the normalized radius $\rho$ = 0 and 0.75. (b) Plasma potential profiles measured by the LHD-HIBP diagnostic at $t = $ 4.0, 6.1, and 7.0 s.} 
    \label{fig:plasma potential profiles}
\end{figure}

Figure~\ref{fig:plasma potential profiles} shows the discharge waveforms and plasma potential profile $\phi$ measured after improving the negative ion beam transport efficiency of the HIBP low-energy beamline. The horizontal axis is the normalized minor radius $r/a$, defined as the effective radius, $r$, divided by the effective minor radius, $a$. The Au$^+$ beam energy injected into the plasma was 4.538 MeV. The plasma potential profile was obtained by sweeping the voltages of beam steering electrodes mounted near the beam injection and energy analyzer ports with a 10 Hz triangular wave. The LHD had the magnetic axis $R_\mathrm{ax}$ = 3.6 m and a magnetic field strength $B_t = 2.75$ T. The central heating of Electron cyclotron heating (ECH) started at $t = 3.0$ s and turned off at $t = 6.0$ s. The neutral beam heating was superimposed at $t = 3.3$ s and remained so until $t = 6.0$ s. The ECH is predominant at $t = 6.0$ s, resulting in a positive plasma potential at $r/a < 0.4$ and a positive radial electric field was formed in 0.1 to 0.4. The plasma potential in the measured region becomes negative at $t = 6.1$ s, with a flat potential profile at $t = 7.0$ s. 

\begin{figure}
    \centering
    \includegraphics[width=0.8\linewidth]{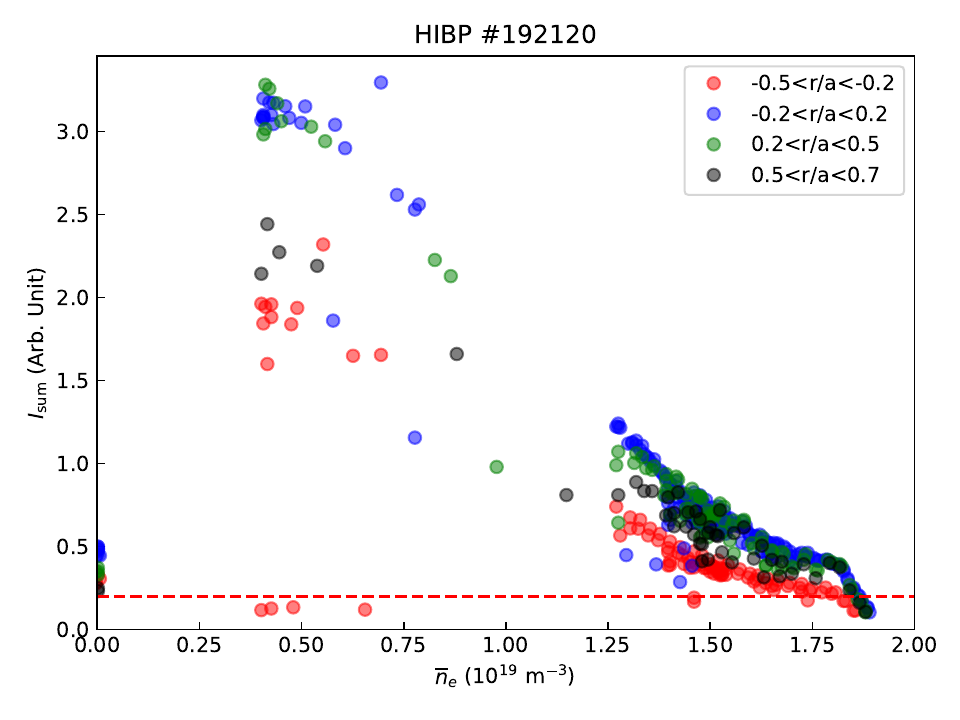}
    \caption{Dependence of the signal intensity $I_{\rm{sum}}$ of the secondary beam Au$^{2+}$ detected by the energy analyzer on the line-averaged density $\overline{n}_{\rm{e}}$ [m$^{-3}$] measured with the interferometer.}
    \label{fig:dependence of secondary beam signal}
\end{figure}

The primary beam undergoes attenuation due to collisions with electrons and ions in the plasma. To evaluate the diagnostic limit with respect to electron density, the relationship between the secondary ion beam signal ($I_{\rm{sum}}$) of Au$^{2+}$ and the line-averaged electron density $\overline{n}{\rm{e}}$ is shown in figure~\ref{fig:dependence of secondary beam signal}. Since the primary beam is swept across the plasma to measure the potential profile, it traverses regions with different local electron densities. Therefore, $I{\rm{sum}}$ is plotted separately for each region of normalized minor radius ($r/a$). The non-zero values of $I_{\rm{sum}}$ observed near $\overline{n}_{\rm{e}} = 0$ arise from collisions between the primary beam and residual neutral particles prior to the plasma discharge, which generate secondary ions. Accordingly, the potential profiles are calibrated shot-by-shot such that the pre-discharge signal corresponds to 0~V. The area below the red dashed line in the figure indicates the noise level during the discharge. In this discharge, although the $I_{\rm{sum}}$ signal in the region $r/a < -0.2$ is slightly lower than in other regions, it remains relatively constant for $r/a > -0.2$. These results indicate that plasma potential measurements are feasible for line-averaged electron densities up to approximately $\overline{n}_{\rm{e}} = 1.75 \times 10^{19}$~m$^{-3}$. At higher densities, however, the signal decreases rapidly, making potential measurements increasingly difficult.

\newpage

\section{Summary}
A new cesium sputter-type negative ion source was introduced into the LHD-HIBP system to enhance the injected beam current and improve the signal quality of plasma potential measurements. Despite the improved extraction current on a test stand, significant beam losses were observed in the low-energy transport section due to space-charge effects, especially for high-current, low-energy Au$^-$ beams. Through IGUN simulations and experimental validation, we demonstrated that these losses can be effectively mitigated by optimizing the electrode voltages in the multistage accelerator to introduce an electrostatic lens effect.

This approach increased the Au$^-$ beam current injected into the tandem accelerator by a factor of approximately 2–3, and the resulting Au$^+$ beam current available for plasma potential diagnostics also increased substantially. Crucially, this optimization was achieved without mechanical alterations to the beamline, relying solely on electronic adjustments. Consequently, plasma potential measurements using HIBP were extended to higher electron density regimes (up to $1.75 \times 10^{19}$~m$^{-3}$) with improved signal integrity.

The demonstrated technique provides a compact, practical, and highly effective approach to improving the transport of heavy-ion beams under strong space-charge conditions. This advancement significantly enhances diagnostic performance and supports the detailed investigation of plasma behavior in magnetically confined fusion devices, thereby contributing to the development of future fusion reactors. In addition to its utility in plasma diagnostics, the method is broadly applicable across diverse accelerator systems, providing a versatile solution to beam transport challenges in both scientific research and industrial applications.

\section*{Acknowledgements}
This work was supported by the JSPS KAKENHI under Grant Nos. 19KK0073 and 23H01160. We thank the LHD experiment group for their generous experimental support and Mr. Y. Nakajima for his support.

The data supporting the findings of this study are available in the LHD experiment data repository at https://doi.org/10.57451/lhd.analyzed-data.

\section*{References}
\NF
\bibliographystyle{iopart-num}
\bibliography{references}

\section*{Author contributions statement}
M.N. and K.N. conceived the simulations and the experiments. M.N., K.N., K.U, A.S, and H.T conducted the experiments and analyzed the results. All authors reviewed the manuscript.
\end{document}